\documentclass[journal=ancac3,manuscript=article,layout=traditional]{achemso}

\usepackage{braket}
\usepackage[version=3]{mhchem}
\usepackage{graphicx}
\usepackage{upgreek} 
\usepackage{bm}

\author{Salman Ahsanullah}
\affiliation{Department of Physics and Astronomy, The University of Kansas, Lawrence, Kansas 66045, United States}

\author{Neema Rafizadeh}
\affiliation{Department of Physics and Astronomy, The University of Kansas, Lawrence, Kansas 66045, United States}

\author{Hui Zhao}
\affiliation{Department of Physics and Astronomy, The University of Kansas, Lawrence, Kansas 66045, United States}
\email{huizhao@ku.edu}

\title[\texttt{achemso} Transient Absorption Spectroscopy of NbOI$_2$]
{Transient Absorption Spectroscopy of NbOI$_2$}


\begin{document}

\begin{tocentry}

  \includegraphics[width=8.5cm]{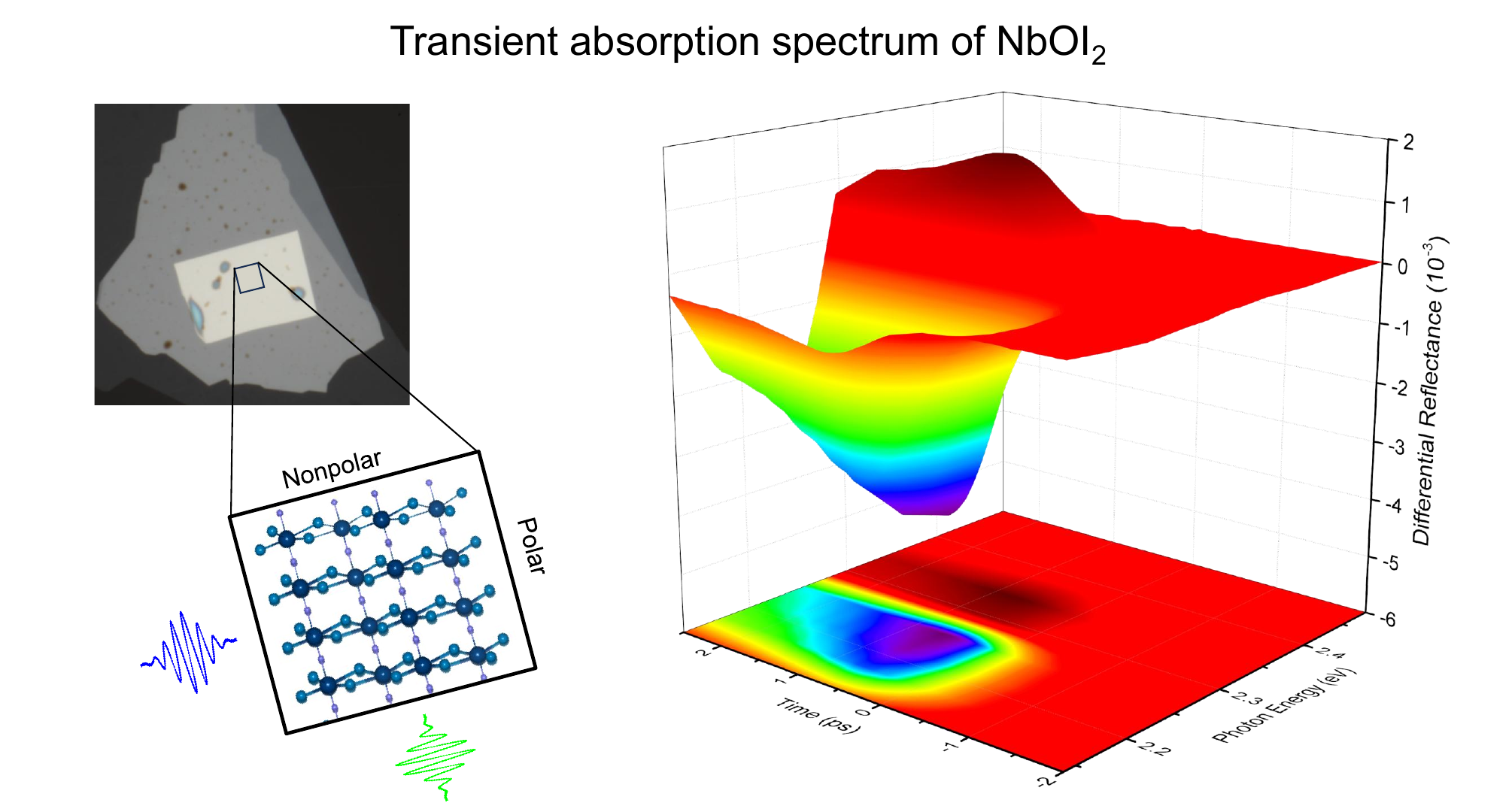}

\end{tocentry}

\begin{abstract}

NbOI$_2$ has recently emerged as a new van der Waals material combining semiconducting behavior with intrinsic in-plane ferroelectricity and pronounced transport and optical anisotropy. However, its photocarrier dynamics remain largely unexplored. Here we report transient absorption spectroscopy of NbOI$_2$ using femtosecond pump–probe measurements. A pronounced transient absorption feature is observed near the 2.34 eV excitonic resonance, arising from photocarrier-induced excitonic energy shifts and saturation. The decay dynamics reveal an exciton lifetime of several tens of picoseconds. The observed density-dependent behavior is consistent with exciton–exciton annihilation and defect-assisted Auger recombination, yielding a rate of 0.4 cm$^2$ s$^{-1}$. Polarization-resolved measurements further reveal a pronounced in-plane anisotropy in the transient response that follows the linear absorption anisotropy. These findings provide fundamental insight into photocarrier dynamics in NbOI$_2$ and establish key parameters for understanding and exploiting its optoelectronic behavior.
\end{abstract}

Layered niobium oxide dihalides NbOX$_2$ (X = Cl, Br, I) have recently emerged as an intriguing family of van der Waals materials that combine semiconducting behavior with intrinsic in-plane ferroelectricity.\cite{np181170,nh41113,np19142,nc164987,small212407729} Among them, NbOI$_2$ has shown particularly promising properties as a prototype two-dimensional (2D) ferroelectric semiconductor. Its low-symmetry monoclinic structure features Nb atoms displaced from the center of [NbO$_2$X$_4$] octahedra, producing spontaneous polarization confined within the layers at room temperature,\cite{am332101505} which enables switchable rectification in double-Schottky-barrier devices\cite{apl119033103} and antiferroelectric ordering.\cite{nh41113,acsnano177170,nc145911}

The Peierls-distorted polar structure of NbOI$_2$, with anisotropic bonding, gives rise to pronounced anisotropy in its structural, electronic, and optical properties.\cite{nanotechnology33275701,jap134085105,nc131884} Mechanical measurements showed a large in-plane anisotropy in Young’s modulus\cite{apl123051905} and directional ferroelectric domains.\cite{npj2dma862,acsnano177170} Angle-resolved Raman and absorption studies revealed strongly directional phonon modes,\cite{am332101505} while anisotropic in-plane electrical conductance and photoresponsivity enabled polarization-sensitive photodetection and synaptic transistor functionalities.\cite{am332101505,am372500049} The polar structure of NbOI$_2$ also leads to exceptional nonlinear optical responses. Multiple studies have reported giant, electrically tunable, and strain-dependent second-harmonic generation in NbOI$_2$, with conversion efficiencies far exceeding those of most 2D materials.\cite{np16644,afm342308207,nl243413,afm2025e16389,apr20252500188,nc145911} NbOI$_2$ further exhibits remarkable piezoelectric and electro-optic properties, with exceptionally large piezoelectric coefficients and substantial linear electro-optic and elasto-optic responses in both 2D and bulk forms.\cite{nc131884,b110L100101,ss42300125} These unique properties have inspired exploration of NbOI$_2$ for various applications such as polarization-sensitive sensors,\cite{ss42300125} nonvolatile and neuromorphic devices,\cite{acsnano1927654,nl2513844} optoelectronic and gas-sensing,\cite{nc164149} and THz emission and detection.\cite{aom132403471,nm241203}

Despite these advances, the semiconducting aspects, particularly the photoexcitation dynamics, of NbOI$_2$ remain largely unexplored. Understanding the photocarrier dynamics of NbOI$_2$ is critical for elucidating the microscopic origins of its extraordinary THz and nonlinear optical responses, and for enabling its use in high-speed ferroelectric optoelectronic and quantum devices. Here, we report time-resolved transient absorption spectroscopy measurements on NbOI$_2$ bulk crystals. We identify a distinct excitonic resonance near 2.34 eV whose transient response reflects both excitonic energy renormalization and saturation effects induced by photoexcited carriers. The temporal evolution reveals a long-lived exciton population persisting for several tens of picoseconds and exhibits clear signatures of exciton–exciton annihilation and defect-assisted Auger recombination. Furthermore, polarization-resolved measurements uncover a pronounced in-plane anisotropy in the transient absorption, consistent with the polar symmetry of the crystal. These results illustrate the experimental picture of exciton relaxation and anisotropic transient response in NbOI$_2$, establishing key parameters for understanding its semiconducting and optoelectronic properties.

 \begin{figure}[t]
  \centering
  \includegraphics[width=15cm]{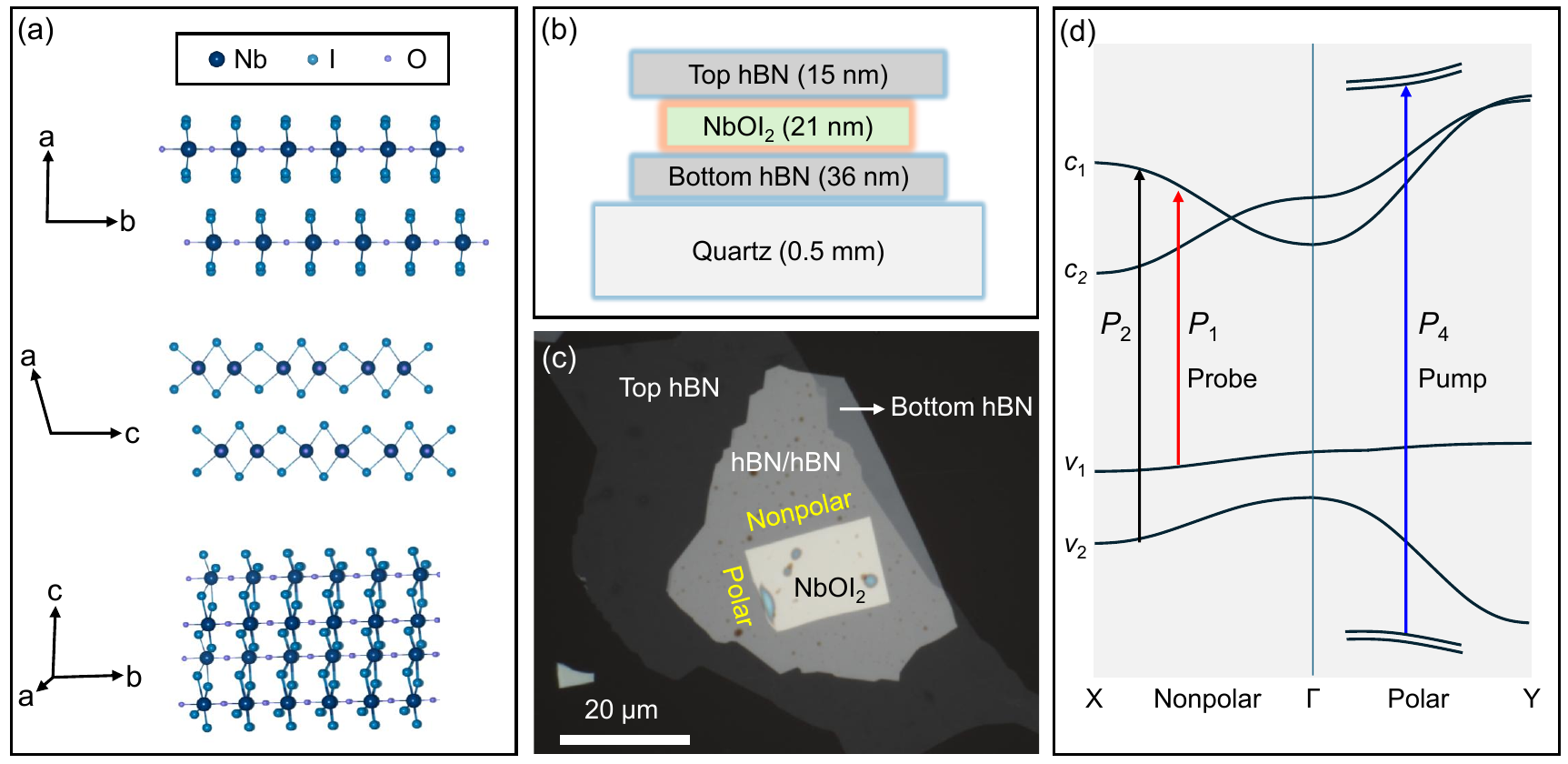}
  \caption{(a) Crystal structure of NbOI$_2$. (b) Device structure showing a NbOI$_2$ flake sandwiched between two hexagonal boron nitride (hBN) flakes on a quartz substrate. (c) Optical microscope image of the device used for optical measurements. (d) Schematic band structure of NbOI$_2$ showing the relevant excitonic transitions.}
    \label{fig:sample}
\end{figure}

Figure \ref{fig:sample}(a) shows the crystal structure of NbOI$_2$. The displacement of Nb atoms along the $b$ axis induces a spontaneous polarization along that direction, which defines the polar axis. We adopted a device structure as shown in Figure \ref{fig:sample}(b), where a NbOI$_2$ flake is sandwiched between two hexagonal boron nitride (hBN) layers on a quartz substrate. The device was fabricated by exfoliating thin flakes of hBN and NbOI$_2$ using adhesive tape and sequentially transferring them onto the substrate, followed by thermal annealing (see Supporting Information). Figure \ref{fig:sample}(c) presents an optical microscope image of the resulting device. The polar and nonpolar edges of the NbOI$_2$ flake is identified according to its anisotropic optical absorption (to be discussed later). Atomic force microscopy measurements (see Supporting Information) show that the NbOI$_2$ flake has a thickness of 21 nm, corresponding to 25 monolayers. The top and bottom hBN layers are 15 nm and 36 nm thick, respectively. The hBN encapsulation is crucial for preventing sample degradation during the optical experiments (see Supporting Information). 

Figure~\ref{fig:sample}(d) schematically illustrates the relevant features of the NbOI$_2$ band structure established by angle-resolved photoemission spectroscopy and density functional theory.\cite{np16644} The conduction band $c_1$ is dominated by Nb $d_{xy}$ character, whereas $c_2$ contains mainly Nb $d_{yz}$
contributions. The topmost valence band $v_1$ originates primarily from Nb $d_{x^2-y^2}$/$d_{z^2}$ orbitals, while $v_2$ derives predominantly from I $p_y$ orbitals. The conduction-band minimum and valence-band maximum are located at the X and Y points, respectively, making NbOI$_2$ an indirect semiconductor. However, ellipsometry and optical absorption measurements have revealed several strong excitonic resonances. The dominant ones in the visible range, at 2.34~eV ($P_1$) and 2.64~eV ($P_2$), arise from the $v_1$--$c_1$ and $v_2$--$c_1$ transitions at \textit{k}-points along the nonpolar direction. In contrast, the $P_4$ (3.54~eV) excitonic resonance originates from transitions between higher-lying conduction and deeper valence bands at \textit{k}-points along the polar direction.\cite{np16644}

Here we focus on the transient absorption of the $P_1$ resonance. We first performed transient absorption spectroscopy measurements in reflection geometry (see Supporting Information). The NbOI$_2$ flake was photoexcited by a 3.02~eV pump pulse with a peak fluence of 13~$\upmu$J~cm$^{-2}$, which is expected to inject exciton by the $P_4$ transition [Figure \ref{fig:sample}(d)]. To estimate the injected carrier density, we measured the reflectance (0.48) and transmittance (0.26) of the sample at this photon energy, which allowed us to deduce an absorption coefficient of $1.43 \times 10^5$~cm$^{-1}$ using the sample thickness of 21 nm. By using Beer's law, we deduced the peak injected areal carrier densities to be $1.7 \times 10^{11}$~cm$^{-2}$ at the first layer of the sample. The change in the sample’s absorption coefficient induced by these carriers (known as transient absorption) was detected by measuring the differential reflectance of a tunable probe pulse near the $P_1$ resonance, defined as $ \Delta R / R_0 = (R - R_0)/ R_0$, where $R$ and $R_0$ are the reflectance of the pump-excited and unexcited NbOI$_2$ device, respectively (Supporting Information).\cite{afm271604509}

 \begin{figure}[t]
  \centering
  \includegraphics[width=10cm]{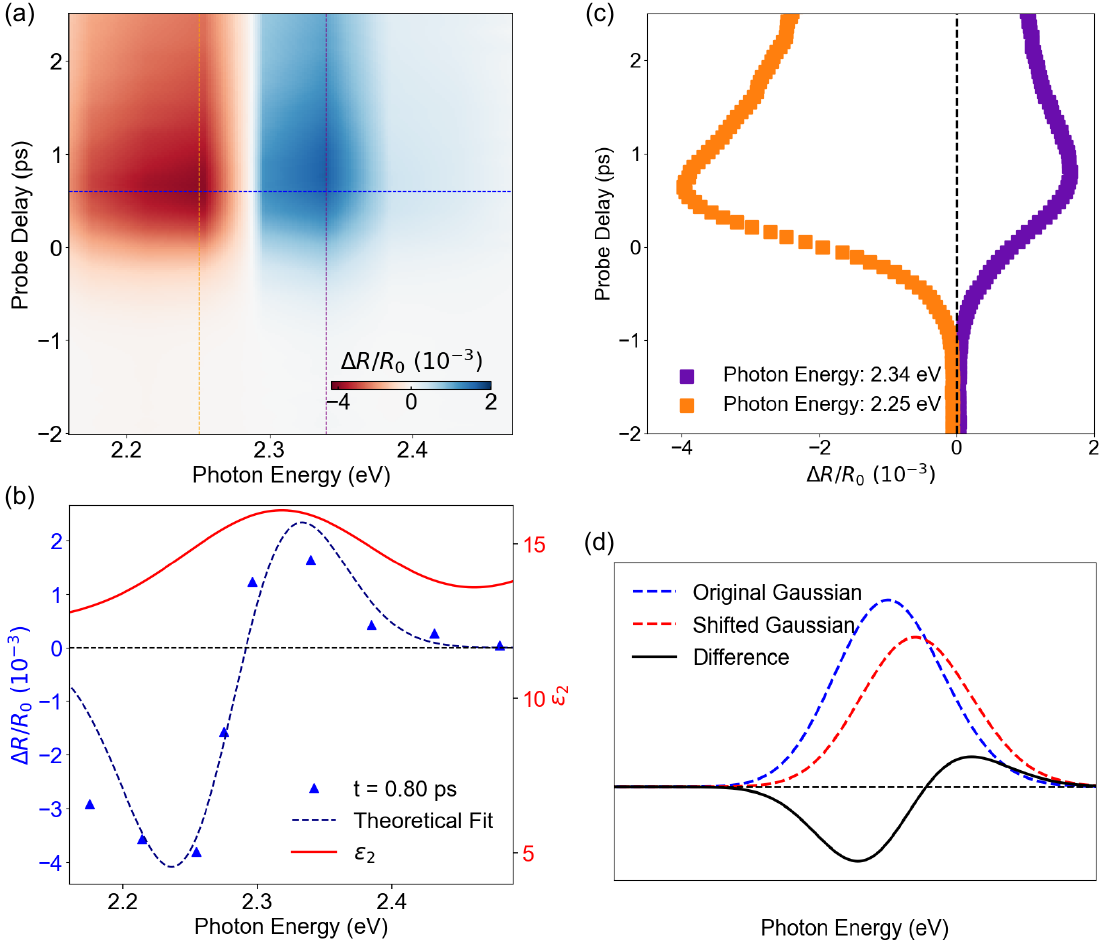}
  \caption{(a) Differential reflectance of NbOI$_2$ as a function of probe delay and probe photon energy. (b) Peak differential reflectance as a function of probe photon energy, corresponding to the dashed horizontal line in (a). The red curve highlights the excitonic resonance, while the dashed curve represents a fit based on a shift and saturation of the exciton resonance. (c) Time-resolved differential reflectance at two probe photon energies corresponding to the positive and negative peaks. (d) Schematic illustration of the origin of the differential reflectance spectrum from a shifted and saturated excitonic absorption peak.}
    \label{fig:TAS}
\end{figure}

Figure~\ref{fig:TAS}(a) shows the differential reflectance signal as a function of probe photon energy and probe delay. In this measurement, the pump and probe pulses are vertically and horizontally polarized, respectively, while the sample polar direction make an arbitrarily selected angle of 75$^\circ$ from horizontal (detailed angular dependence will be discussed later). A pronounced spectroscopic feature near the 2.34~eV resonance with a fast temporal evolution is observed. The peak signal, obtained at a delay of 0.8~ps, is plotted in Figure~\ref{fig:TAS}(b) as triangles, together with the absorption lineshape of this resonance (red curve) re-plotted from Ref.~\citenum{np16644}. The rapid rise of the signal, as shown in Figure~\ref{fig:TAS}(c), provides unambiguous evidence that the transient absorption originates from pump-injected photocarriers. It is well established that photocarriers in semiconductors can induce transient absorption near an excitonic resonance through energy shift, saturation, and broadening of the transition.\cite{b326601,n411549} By using a blue shift of 0.25~meV and a saturation of 0.14\%, we can reproduce the observed spectral shape [dashed curve in Figure~\ref{fig:TAS}(b)], as schematically illustrated in Figure~\ref{fig:TAS}(d).

 \begin{figure}[t]
  \centering
  \includegraphics[width=10cm]{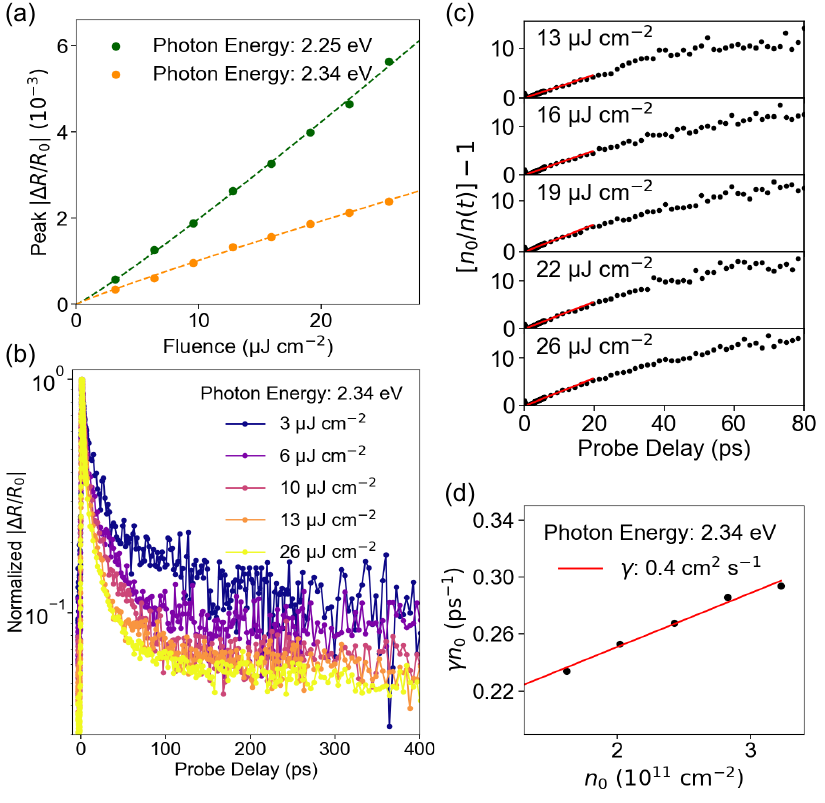}
  \caption{(a) Peak differential reflectance magnitude of NbOI$_2$ as a function of pump fluence, measured with probe photon energies of 2.25 and 2.34 eV, respectively. (b) Representative time-resolved differential reflectance traces at various pump fluences, measured with a 2.34 eV probe. (c) ${n_0}/{n(t)} - 1$ calculated from the data as a function of probe delay for several high pump fluences. The red lines represent linear fits to the data at early delays. (d) $\gamma n_0$ extracted from the fits shown in (c) as a function of the injected carrier density $n_0$. The red line indicates a linear fit.}
    \label{fig:power}
\end{figure}

We next examined how the signals at the two extreme probe photon energies vary with pump fluence. Under linear absorption, the pump fluence is proportional to the injected carrier density. Although NbOI$_2$ possesses relatively large second- and third-order nonlinear susceptibilities, \cite{np16644,apr20252500188} our experiment is conducted at low peak irradiance on the order of $10^8$~W~cm$^{-2}$. At this irradiance, carrier injection via two-photon absorption or absorption of the second-harmonic light is negligible (see Supporting Information). Figure~\ref{fig:power}(a) shows that the peak signal magnitude increases linearly with pump fluence. This linearity establishes that the differential reflectance is proportional to the carrier density and therefore accurately tracks the carrier dynamics.

Owing to the indirect bandgap of NbOI$_2$, the quasi–steady-state photocarrier population is expected to consist mainly of electrons near the X point and holes near the Y point
[Fig.~\ref{fig:sample}(d)]. We hypothesize that the dominant photoexcited species are excitons formed by these electrons and holes. Giant exciton binding energies on the order of 600--800~meV have been reported in bulk NbOCl$_2$, a material with the same lattice structure and similar electronic
properties.\cite{n61353,jpcl159191} Although the exciton binding energy of NbOI$_2$ has not yet been determined, it is reasonable to assume that it is well above 26~meV, making the excitons stable at room temperature. The normalized time-resolved differential reflectance at various pump fluences are
shown in Fig.~\ref{fig:power}(b), which clearly reveals two features. First, the signal persists for several tens of picoseconds at all fluences, reflecting the exciton lifetime in NbOI$_2$. Second, the decay becomes faster at higher fluence, indicating density-dependent excitonic dynamics.

Previously, exciton--exciton annihilation (XXA) and defect-assisted (DA) Auger recombination have been established as dominant nonlinear exciton decay channels in layered semiconductors at elevated densities, and both scale quadratically with exciton density. In the former, two excitons interact and one recombines nonradiatively, transferring its energy to the other.\cite{b89125427,nl145625,l124057403} In the latter, a defect traps an exciton (or one of its carriers), and the trapped species recombines with a free exciton. The presence of defect states relaxes crystalline momentum conservation constraints, allowing Auger-like decay to occur efficiently even at moderate densities.\cite{nl15339,b91165411,afm264733} With such quadratic decay channels included, the rate equation for the exciton density ($n$) can be written as
\begin{equation}
\frac{dn}{dt} = -\frac{n}{\tau} - \frac{1}{2}\gamma n^{2},
\label{eq:rate}
\end{equation}
where $\tau$ denotes the single-particle exciton lifetime. We define $\gamma = \gamma_{\mathrm{XXA}} + \gamma_{\mathrm{DA}}$, where $\gamma_{\mathrm{XXA}}$ and $\gamma_{\mathrm{DA}}$ are the bimolecular coefficients for XXA and DA Auger recombination, respectively. When these nonlinear decay channels dominate the exciton dynamics, the first term on the right-hand side of Eq.~\ref{eq:rate} can be neglected, leading to a simple solution:
\begin{equation}
\frac{n_{0}}{n(t)} - 1 = \gamma n_{0} t,
\label{eq:XX}
\end{equation}
where $n_{0}$ is the initially injected exciton density at $t=0$.

To compare our data with this model, we calculated ${n_0}/{n(t)} - 1$ from the time-resolved differential reflectance at elevated pump fluences using the linear relation established in Figure~\ref{fig:power}(a), and plotted the results in Figure~\ref{fig:power}(c). At early delays, this quantity increases approximately linearly, as expected from the model. By performing linear fits (red lines), we extracted the parameter $\gamma n_0$ as a function of $n_0$, as shown in Figure~\ref{fig:power}(d). The resulting linear dependence further confirms the validity of this model for early probe delays and yields a rate of $\gamma = 0.4 \pm 0.1~\text{cm}^{2}\,\text{s}^{-1}$. This value is on the same order of magnitude with previously reported exciton-exciton annihilation rates in 2D transient metal dichalcogenides.\cite{b89125427,nl145625} However, our analysis cannot conclusively separate the contributions of the two nonlinear decay mechanisms. We further note that such decay channels are insignificant on pulsewidth-limited ultrafast time scales, as evidenced by the linear dependence of the peak signal on pump fluence [Figure~\ref{fig:power}(a)].

Finally, we explore the in-plane anisotropy of the transient absorption. The wavefunction of the $P_1$ exciton is confined along the nonpolar direction, \cite{np16644} and therefore primarily interacts with light polarized along this direction through dipole coupling. In contrast, the higher-energy $P_4$ exciton is confined along the polar direction and thus couples most strongly to light polarized along the polar axis. We first measured the polarization-dependent linear absorption of the sample at the $P_1$ and $P_4$ resonances using the probe and pump pulses individually. As shown in Figure~S2, the absorptance at $P_1$ reaches a maximum when the light polarization is aligned with the nonpolar direction, whereas the absorptance at $P_4$ peaks when the polarization is aligned with the polar direction of NbOI$_2$. These behaviors are consistent with previous report.\cite{np16644}

 \begin{figure}[t]
  \centering
  \includegraphics[width=10cm]{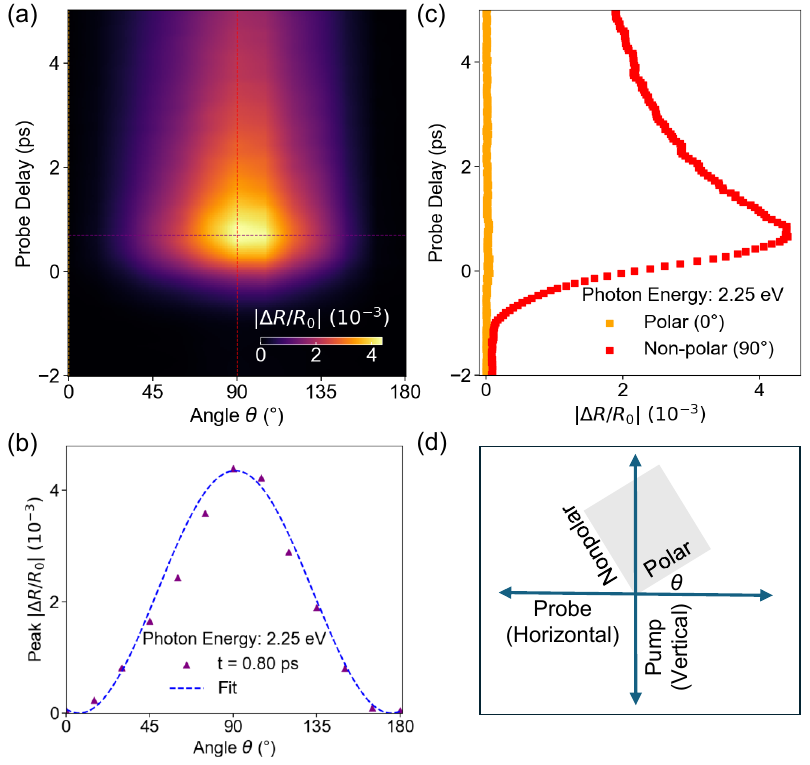}
  \caption{(a) Differential reflectance of NbOI$_2$ as a function of sample orientation and probe delay. (b) Peak differential reflectance as a function of the angle between the crystal’s polar direction and the probe polarization. The dashed line represents a fit. (c) Time-resolved differential reflectance at selected sample orientations. (d) Measurement configuration showing the pump and probe polarizations (vertical and horizontal in the laboratory frame, respectively) and the definition of the angle $\theta$. }
    \label{fig:angle}
\end{figure}

Figure~\ref{fig:angle}(a) shows the measured time-resolved differential reflectance at various sample orientations. In these measurements, the pump and probe pulses are linearly polarized along the horizontal and vertical directions in the laboratory frame, while the sample is rotated about its surface normal. This rotation varies the angle $\theta$ between the polar axis of the sample and the probe polarization, as illustrated in Figure~\ref{fig:angle}(d). The signal at 0.8~ps is plotted in Figure \ref{fig:angle}(b) as a function of $\theta$, which shows a characteristic $\mathrm{sin}^2(\theta)$ dependence (dashed curve). The time scans at the two extreme angles are shown in Figure \ref{fig:angle}(c). The strong angular dependence of the transient absorption confirms the orthogonal confinement of the $P_1$ and $P_4$ excitons.\cite{np16644}

In summary, our transient absorption measurements revealed key aspects of the photocarrier dynamics in NbOI$_2$. The excitonic resonance near 2.34 eV exhibits pronounced transient absorption due to photocarrier-induced energy renormalization and saturation effects. We deduced an exciton lifetime on the order of several 10's of picosecond. The observed density-dependent exciton recombination is consistent with exciton–exciton annihilation and defect-assisted Auger recombination. The in-plane anisotropy of the transient response reflects the intrinsic polar symmetry of NbOI$_2$ and its strongly directional optical coupling. These findings reveal the microscopic processes governing exciton relaxation and interaction in this emerging ferroelectric semiconductor and provide a foundation for developing ultrafast, polarization-sensitive optoelectronic and quantum devices based on van der Waals ferroelectric semiconductors.

\begin{acknowledgement}

This work was supported by the U.S. Department of Energy, Office of Science, Basic Energy Sciences, under Award \# DE-SC0020995.

\end{acknowledgement}

\begin{suppinfo}
Sample fabrication and characterization, linear absorption spectroscopy data, analysis of nonlinear absorption mechanisms, transient absorption spectroscopy setup.

\end{suppinfo}

%\bibliography{reference}

\providecommand{\latin}[1]{#1}
\makeatletter
\providecommand{\doi}
  {\begingroup\let\do\@makeother\dospecials
  \catcode`\{=1 \catcode`\}=2 \doi@aux}
\providecommand{\doi@aux}[1]{\endgroup\texttt{#1}}
\makeatother
\providecommand*\mcitethebibliography{\thebibliography}
\csname @ifundefined\endcsname{endmcitethebibliography}  {\let\endmcitethebibliography\endthebibliography}{}

\end{document}